\documentclass[twocolumn,superscriptaddress,prapplied,bibliography]{revtex4-1}
\usepackage{graphicx}
\usepackage{amssymb}

\usepackage{amsmath}
\usepackage{epsfig}
\usepackage{color}
\usepackage{mathtools}

\begin{document}
\title{High resolution imaging with anomalous saturated excitation}

\author{Bo Du}
\author{Xiang-Dong Chen}
\author{Shao-Chun Zhang}
\author{Ze-Hao Wang}
\author{En-Hui Wang}
\author{Guang-Can Guo}
\author{Fang-Wen Sun}
\email{xdch@ustc.edu.cn}

\affiliation{CAS Key Lab of Quantum Information, University of Science and Technology of China, Hefei, 230026, P.R. China}
\affiliation{CAS Center For Excellence in Quantum Information and Quantum Physics, University of Science and Technology of China, Hefei, 230026, P.R. China}

\begin{abstract}
The nonlinear fluorescence emission has been widely applied for the high spatial resolution optical imaging. Here, we studied the fluorescence anomalous saturating effect of the nitrogen vacancy defect in diamond. The fluorescence reduction was observed with high power laser excitation. It increased the nonlinearity of the fluorescence emission, and changed the spatial frequency distribution of the fluorescence image. We used a differential excitation protocol to extract the high spatial frequency information. By modulating the excitation laser's power, the spatial resolution of imaging was improved approximate 1.6 times in comparison with the confocal microscopy. Due to the simplicity of the experimental setup and data processing, we expect this method can be used for improving the spatial resolution of sensing and biological labeling with the defects in solids.
\end{abstract}

\maketitle

\section{Introduction}
With stable photon emission and optically controlled spin, the fluorescent defects in solids have been widely studied for the quantum information processing, sensing and biological imaging \cite{wra2006review,Doherty2013phyrrevie,degen-review-2014,siv-prl2,Dyakonov-hbnspin-2020nm,Akimov-gev-2018acsp,Awschalom-sic-sc2020}. One of the most promising candidates is the nitrogen vacancy (NV) center in diamond. Due to its atomic size, the sensing and imaging with NV center show the advantage of high spatial resolution. Several optical super-resolution microscopy techniques have been demonstrated for the sub-diffraction resolution imaging of NV center\cite{hell-2009,xip-sim,wra2014PNASstorm,chen201501,englund-resolution-npjQI2019,Walsworth-spinresolft-2017oe,gum-nanoscopy-lsa2017}. Shape-modulation of the laser beam or selective excitation of single NV centers is usually required for these imaging methods.

Meanwhile, the nonlinear optical response of the fluorescence is also explored to improve the spatial resolution of the microscopy \cite{Gustafsson-nsim-2012pnas,superlinear19,xuliu-prl2018,Heintzmann-satu02,Chen2018OL,harmonics2017}. The key is to increase the high spatial frequency component of the image, and suppress the low spatial frequency component. For example, saturated excitation microscopy (SAX) has been demonstrated by utilizing the nonlinearity of the saturation\cite{fujita07prl,fujita18} . The high spatial frequency component of a fluorescence image can be extracted with the harmonic demodulation or the differential excitation\cite{kuang-2020lpr-review}.
The nonlinearity of the fluorescence determines the resolution and signal-to-noise ratio of the images.
Therefore, emitters with a high nonlinearity of the fluorescence emission are required for the high resolution imaging \cite{tinnefeld10,sauer09,superlinear19}. Combined with STED microscopy, the nonlinear fluorescence emission has also been utilized for the super-resolution imaging with upconversion nanoparticles\cite{Ploschner-oe2020-sted,simone-2020nanoscale,chench-2018nc}.

Recent experiments reveal that the excitation dependence of the NV center fluorescence is different from a traditional saturating\cite{chen20152prb,chapman-anosatu-12,njp-darkstate}. Fluorescence reduction is observed with a high-intensity laser excitation. This effect could enhance the nonlinearity of fluorescence emission. Here, by utilizing this anomalous saturation effect, we demonstrated a sub-diffraction microscopy for the NV center imaging.
A single Gaussian-shaped laser beam pumped the fluorescence of NV center through a scanning confocal microscope. Due to the fluorescence reduction with a high intensity laser excitation, a doughnut-shaped fluorescence image with high spatial frequency component was obtained. To extract the high spatial frequency information of the images, we applied a differential excitation protocol, where the fluorescence intensities of NV center with high and low power lasers were simultaneously recorded and compared. The results showed that the resolution of NV center imaging can be improved 1.6 times by utilizing the anomalous saturating. With the simple experimental setup and data processing, we expected this anomalous saturated excitation (ASAX) microscopy can be used to improve the spatial resolution of NV center's sensing and imaging.

\begin{figure*}[hbtp]
\centering\includegraphics[width=12cm]{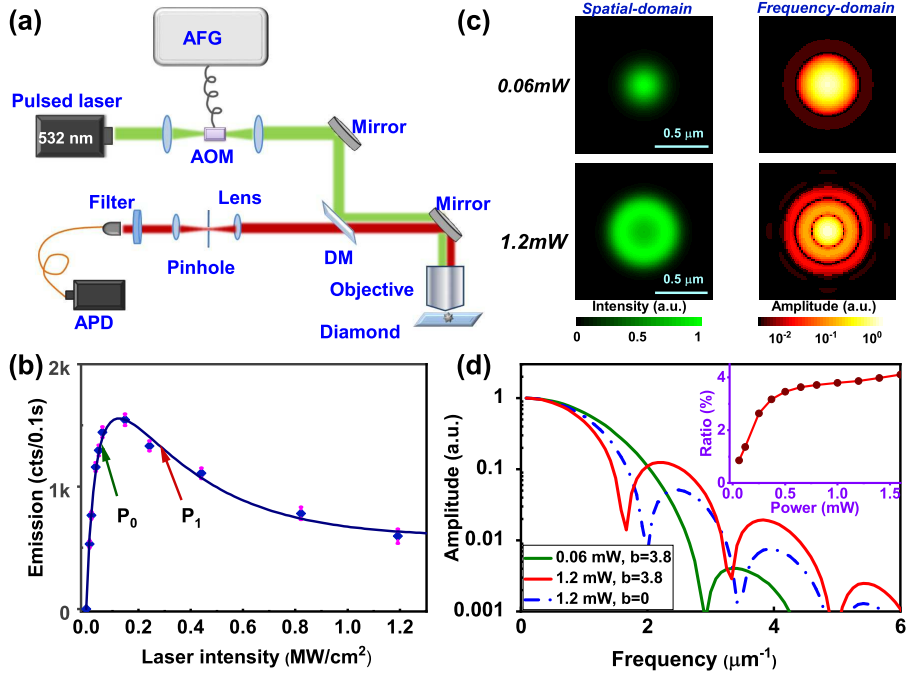}
\caption{(a) Experimental setup. The pulsed 532 nm laser is circularly polarized, with a repetition rate of 5 MHz. DM: dichroic mirror; APD: avalanche photodiode. (b) The fluorescence emission of a single NV center with different laser intensities. The blue line is the fitting with Eq.\ref{eqsatu}. (c) Simulated point spread function and the spatial frequency distribution of the confocal microscopy with different powers. (d) The normalized spatial frequency with the excitation power of 0.06 and 1.2 mW. For comparison, we also show the results with a traditional saturating (b = 0 in Eq.\ref{eqsatu}). The insert is the ratio of the high spatial frequency signal with different excitation laser powers.} \label{figprinciple}
\end{figure*}

\section{Principle and experimental setup}

The NV center in diamond usually shows two charge states: the negatively charged NV$^{-}$ and the neutrally charged NV$^{0}$. The ground state and excited state of NV$^{-}$ are spin triplet. The excited state of NV$^{-}$ can decay to the ground state through the spontaneous emission or the non-radiative inter-system crossing (ISC). The ISC includes the decay from the excited state to a metastable state, and the subsequent decay from the metastable state to the ground state. The transition between the excited state and the metastable state mainly occurs with $m_{s} = \pm 1$ in the excited state. And the NV center at the metastable state prefers to decay to the ground state with electron spin $m_{s}=0$.
As a result, the fluorescence intensity of NV$^{-}$ with electron spin $m_{s}=0$ is higher than that with $m_{s}= \pm 1$. The ISC process also polarizes the spin state to $m_{s} = 0$. The property of NV$^{-}$ electron spin enables it to be widely used as a qubit, while the spin of NV$^{0}$ is rarely studied. Therefore, 'NV center' in this work, unless otherwise specified, will refer to the NV$^{-}$ charge state. The zero phonon line of NV$^{-}$ spontaneous emission is at 637 nm, while the phonon sideband with a peak around 700 nm is observed at room temperature.

The charge state conversion between NV$^{-}$ and NV$^{0}$ can be pumped by laser with ultra-violet and visible wavelength. It has been demonstrated that the charge state conversion is a spin depolarization process\cite{chen20152prb}, which will decrease the fluorescence intensity of NV center. The fluorescence reduction during the photon-induced spin depolarization process can be written as $1+b\cdot e^{-\gamma t}$. $\gamma$ is the charge state conversion rate, and $t$ is the laser duration. $b$ presents the amplitude of fluorescence reduction, which is determined by the fluorescence difference between $m_{s} =0$ and $m_{s}= \pm 1$ and the spin state initialization fidelity. For a high power laser excitation, the charge state conversion rate will linearly increase with the excitation laser power $\gamma \propto P$\cite{chen-prapp-2017,ioni-arxiv-2012}.  Therefore, the fluorescence reduction that is induced by the spin depolarization can be written as $1+b\cdot e^{-P/P_{1}}$, where $P_{1}$ is the saturation power of this spin depolarization effect. Here, we assume that the width of laser pulse does not change with the laser intensity.

\begin{figure*}[hbtp]
\centering\includegraphics[width=11cm]{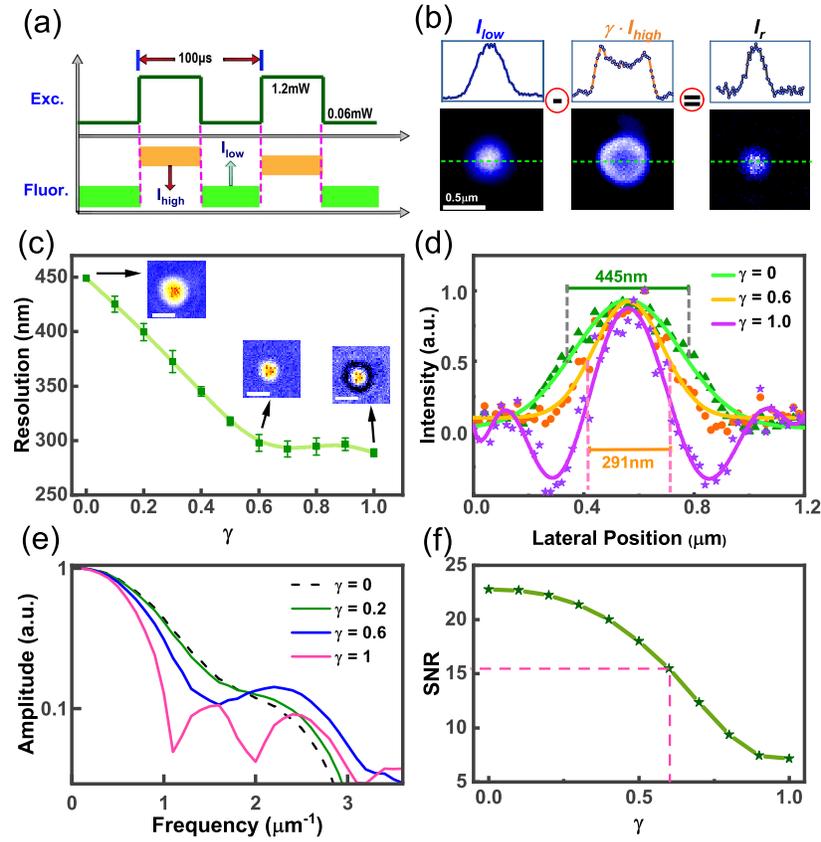}
  \caption{(a) The  sequences of the laser's modulation and fluorescence detection. (b) The images of a single NV center. The confocal microscopy images of $I_{low}$ and $I_{high}$ are recorded to calculate the final image $I_{r}$. Cross-section profiles corresponding to the dashed lines in the images of single NV center are shown on the upsides. (c) The spatial resolution of ASAX microscopy with different $\gamma$. The error bars are the deviation of the results. The insets are images of single NV center with $\gamma$=0, 0.6 and 1.0. The scale bars are 500 nm in length. (d) The lateral cross-section profiles of ASAX images with $\gamma$=0, 0.6 and 1.0. (e) The spatial frequency distribution of ASAX images with different $\gamma$. (f) The SNR of the $I_r$ as a function of $\gamma$. }\label{figsingle}
\end{figure*}

Based on the theoretical analyzing, we can write the excitation power dependence of NV center fluorescence by including the traditional saturating and the spin depolarization:
\begin{equation}
I(P) =I_{sat} \frac{P}{P+P_0}(1+b\cdot e^{-P/P_{1}}). \label{eqsatu}
\end{equation}
Here, $\frac{P}{P+P_0}$ presents the saturation of the ground state to excited state transition, where $P_{0}$ is the saturation power of this transition. $I_{sat}$ is the fluorescence intensity with an infinity excitation intensity.

\begin{figure*}
  \centering
  \includegraphics[width=13cm]{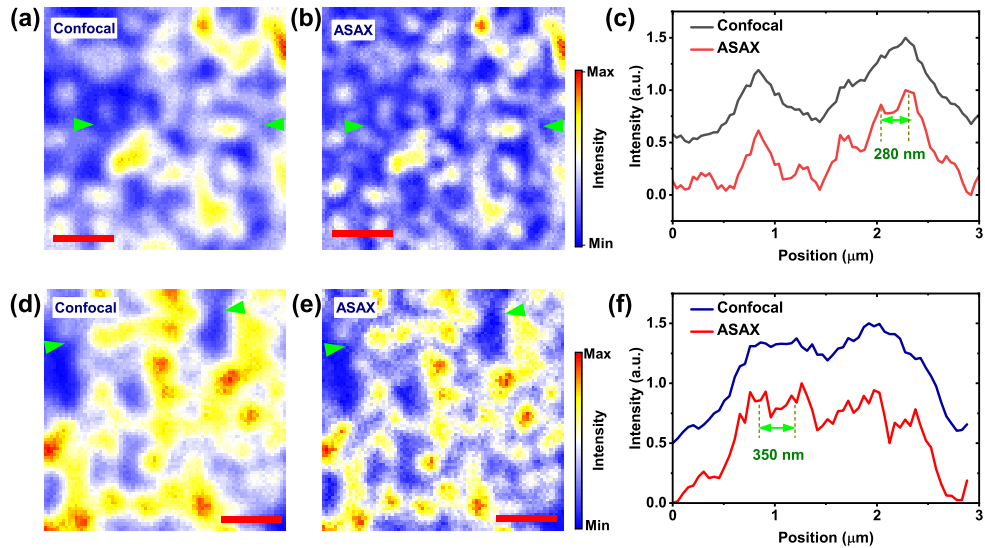}\\
  \caption{ Multiple NV centers imaged with the confocal (a)(d) and ASAX microscopies (b)(e). The scale bars are 1 $\mu m$ in length. (c) and (f) are the cross-section profiles indicated by the arrows.}\label{figdouble}
\end{figure*}

For sub-diffraction NV center imaging, the experiments were carried out with a home-built confocal microscope, as shown in Fig. \ref{figprinciple}(a). NV centers in a chemical vapor deposition diamond plate were produced by nitrogen ion implanting and subsequent annealing. A 0.9 N.A objective was used to focus the laser beam and collect the fluorescence of NV center. The laser beam was a 532 nm picosecond pulsed laser with a Gaussian shape. The laser passed through an acousto-optic modulator (AOM), the diffraction efficiency of which was modulated by an external analog signal (0 - 1 V) from an arbitrary function generator (AFG). The first order of the diffraction beam was used to excite the NV center. In this way,  we can change the power of the excitation laser at a time scale of nanoseconds. A long pass dichroic mirror with an edge wavelength of 640 nm separated the excitation laser and the fluorescence. A long pass optical filter with an edge wavelength of 659 nm was used to further block the excitation laser and the spontaneous emission from NV$^{0}$. Finally, the phonon sideband of NV$^{-}$ center fluorescence was detected by an avalanche photodiode. And the numbers of photon counts were recorded by a data acquisition card for further analysis.

\section{Results}

The fluorescence emission of a single NV center under the excitation of a pulsed laser with a repetition rate of 5 MHz was measured in Fig. \ref{figprinciple}(b). As expected, it showed that the fluorescence intensity increased with the excitation power in the weak pumping region, but decreased with the laser power in the strong pumping region. The thermal effect was excluded by measuring the spin transition signal of NV center\cite{chen20152prb}. The results in Fig. \ref{figprinciple}(b) can be well fitted by Eq. \ref{eqsatu}, with the coefficients of $P_{0}$ = 0.048 MW/cm$^{2}$ and $P_{1}$ = 0.32 MW/cm$^{2}$. Here, with a 5 MHz repetition rate of the excitation laser, the highest fluorescence intensity was observed with an average laser intensity of 0.128 MW/cm$^2$. The contrast between the highest fluorescence intensity and the saturated fluorescence intensity was $\frac{I_{max}-I_{sat}}{I_{max}}=0.61$,  corresponding to a fluorescence reduction amplitude of $b \approx 3.8$.

Experiments indicate that the contrast of fluorescence reduction will change with the repetition rate of the pulsed laser\cite{chen20152prb,njp-darkstate}. It is due to  the spin polarization effect during the ISC, which occurs in a time of approximate 200 ns without the excitation of laser. For the pulsed laser with a repetition rate higher than 5 MHz, the spin polarization probability will increase with the interval between two laser pulses. After the spin is initialized, the laser pulse will depolarize the spin through charge state conversion. Therefore, the amplitude of fluorescence reduction will decrease with the repetition rate of excitation laser.\\

The anomalous saturation effect increases the nonlinear signal of fluorescence emission with a high power excitation. It will change the spatial frequency of the confocal microscopy image. In Fig. \ref{figprinciple}(c), we simulated the fluorescence intensity and spatial frequency distribution of the confocal microscopy by using the experimental results of Fig. \ref{figprinciple}(b).  The powers of the Gaussian-shaped laser beam were set to 0.06 mW and 1.2 mW, corresponding to the intensity of 0.013 MW/cm$^2$ and 0.256 MW/cm$^2$ at the beam center,  respectively. It showed that, with a high laser power, the image of a scanning confocal microscope changed to a doughnut shape. In Fig. \ref{figprinciple}(d), we showed the spatial frequency distribution with different powers. To quantitatively present the relation between spatial frequency distribution and the excitation laser power, we calculated the ratio of spectral component with the spatial frequency higher than that corresponded to the resolution of the confocal microscopy. As shown in the insert of Fig. \ref{figprinciple}(d), the component of the high spatial frequency was improved by increasing the laser power. The spatial frequency distribution of the traditional saturating was also simulated by setting $b=0$ in Eq. \ref{eqsatu}. It showed that the high spatial frequency component with anomalous saturation was higher than that with the traditional saturation.

To experimentally obtain a sub-diffraction resolution image, a differential excitation protocol was applied. The key is to simultaneously record two or more images with different spatial frequency distributions. It can be realized by changing the shape of excitation laser beam\cite{Zhao-17optica,xuliu-prl2018,Kseniya-16,sub91}. Here, we modulated the power of the excitation laser to obtain images with different spatial frequency distributions. Two excitation laser powers were used. To ensure a large spatial frequency difference between two images, the peak intensity of the laser beam for the weak excitation was lower than $P_{0}$. And for the strong excitation, the peak intensity of the laser beam was at the scale of $P_{1}$. A square wave pulse sequence with a period of 100 $\mu s$ was generated by the AFG to control the diffraction efficiency of AOM. As shown in Fig. \ref{figsingle}(a), The high and low levels of the square wave sequence were 1 and 0.2 V, respectively. And the duty ratio of the pulse sequence was 50\%. Then the excitation laser power would switch between P and $\frac{1}{20}$P with a period of 100 $\mu s$. The fluorescence intensities with the high and low power excitation were separately recorded by two channels as $I_{low}$ and $I_{high}$.\\

In Fig. \ref{figsingle}(b), we showed the image of a single NV center. The high and low levels of laser power were set to 1.2 mW and 0.06 mW. The scanning confocal image was obtained with a pixel dwell time of 50 ms. As expected the fluorescence image of $I_{high}$ showed a doughnut shape, while the image of $I_{low}$ was Gaussian shape. To obtain the image with a sub-diffraction spatial resolution, we simply calculated a signal as
 \begin{equation}\label{subtraction}
 I_{r}=I_{low}-\gamma\times I_{high}.
\end{equation}
Here, we have normalized $I_{low}$ and $I_{high}$ by dividing the maximum of each channel. An intuitive understanding of this calculation is that the off center signal of $I_{low}$ is suppressed because of the doughnut shape of the image with $I_{high}$. It would subsequently improve the spatial resolution of imaging with $I_{r}$. With a small $\gamma$, the image of $I_{r}$ is almost the same as $I_{low}$. And with a large $\gamma$, $I_{r}$ would be mainly determined by the distribution of $I_{high}$. By adjusting the factor $\gamma$, the suppression of low spatial frequency signals will be optimized.

In Fig. \ref{figsingle}(c), we presented the single NV center imaging with different $\gamma$. For $\gamma < 0.6$, the image of $I_{r}$ still showed a single peak. The full width at half maximum (FWHM) of the image decreased with the factor $\gamma$. For $\gamma > 0.6$, negative values would emerge for $I_{r}$. The image can not be simply fitted by a single peak function. And the width of the image did not show a further decrease by increasing the factor $\gamma$. Specifically, in Fig. \ref{figsingle}(c), the FWHM of the image $I_{r}$ with $\gamma = 0.6$ was 291 nm, while the FWHM of the confocal image ($\gamma = 0$) was 445 nm, as shown in Fig. \ref{figsingle}(d).  It demonstrated that the resolution of imaging can be improved 1.6 times by utilizing the ASAX microscopy.
The frequency distributions with different $\gamma$ were also shown in Fig. \ref{figsingle}(e). The results indicated that the ratio of the high-frequency component reached its maximum at $\gamma=0.6$. It agreed with the change of spatial resolution with $\gamma$. On the other hand, the signal-to-noise ratio (SNR) determines the visibility of the image. By increasing $\gamma$, the signal (the peak-to-peak value of fluorescence image) will decrease, while the noise is increased. For the single NV center imaging in Fig. \ref{figsingle}(f), it showed that the SNR decreased from 23 to 15 by increasing $\gamma$ from 0 to 0.6. And it further decreased to approximate 7 with $\gamma =1$. Therefore, considering both the spatial resolution and SNR, the subtractive factor $\gamma$ was chosen to be 0.6 for the best imaging quality in our system. The noise mainly originated from the shot noise. It caused the fluctuation of photon counts at each pixel. For the imaging in this work, we chose the pixel size much smaller than the FWHM of the point spreading function. Therefore, the shot noise would mainly affect the high spatial frequency distribution. It would not significantly affect the estimation of spatial resolution. But the noise might decrease the accuracy of NV center localization and the contrast of the image.

To further confirm the improvement of the spatial resolution with ASAX microscopy, we applied this technique for the imaging of multi NV centers. In Fig.\ref{figdouble},  the high (low) level of the excitation laser power was set to 1.4 (0.07) mW. To extract the high spatial frequency signal, the factor $\gamma$ was set to 0.6. We compared the images of confocal microscopy ($\gamma = 0$) and ASAX microscopy. It confirmed that the resolution for high density NV center imaging can be improved with ASAX microscopy. As indicated in the cross-section profiles (Fig.\ref{figdouble}(c) and (f)), the NV centers with distances of 280 and 350 nm can be hardly distinguished with confocal microscopy, as the distances were shorter than the resolution of confocal microscopy. In contrast, the image of ASAX microscopy separated the signal from multi NV centers. However, for the imaging with high density NV center, the signal from adjacent emitters contributed to the background of image with high excitation power. It might decrease the signal to noise ratio of the final image with differential excitation\cite{chench-2018nc}.

\section{Discussion and conclusion}

Solid state defect with spin-dependent fluorescence emission is of interest to researches of quantum information processing, biological imaging and nanophotonics. And the charge state conversion has been observed in various solid state defects\cite{Meriles-prl2018-sivcsc,Awschalom-cscsic-2017nc}. We expected that the fluorescence modulation with charge state manipulation in these defects could be utilized to develop the super resolution imaging technique. It could help to realize the high resolution spin state manipulation and detection. The ASAX microscopy provided one of the most simple techniques for detecting the defect with a resolution below the diffraction limit. It can be applied for the high resolution imaging of fluorescent defects and other particles with similar nonlinear optical response\cite{Lixp-APLp2018-ssc}. According to the results, the amplitude of fluorescence reduction with the high power excitation determines the final imaging quality. Since the spin depolarization of NV center changes the fluorescence lifetime of NV center, the time-gated fluorescence detection technique \cite{chen-prappl2019-time,AO20} could be used to improve the amplitude of fluorescence reduction. Other methods, such as that using laser beams with different shapes can also be applied to further improve the spatial resolution of the microscopy with anomalous saturated excitation\cite{kuang-16oe-fed}.

In summary, we studied the anomalous saturation of NV center fluorescence emission. The fluorescence reduction with high power excitation increased the nonlinearity of fluorescence emission. The sub-diffraction ASAX microscopy was demonstrated by extracting the high spatial frequency information with the high power laser excitation. The spatial resolution of NV center imaging was improved approximately 1.6 times in comparison with the confocal microscopy. Considering the simple experimental setup and data processing of this microscopy method, we expect that the technique can be applied for the quantum sensing and biological imaging with NV center and other defects.

\section*{Funding}
This work was supported by National Key Research and
Development Program of China (No. 2017YFA0304504); Anhui Initiative in Quantum Information Technologies (AHY130100); National Natural
Science Foundation of China (Nos. 91536219, 91850102).

\section*{Disclosures}

The authors declare no conflicts of interest.


%

\end{document}